\documentclass[pdflatex,iicol,sn-mathphys-num]{sn-jnl}
\usepackage[T1]{fontenc}
\usepackage[english]{babel}
\addto\captionsenglish{\renewcommand{\figurename}{Fig.}}
\usepackage{graphicx,dcolumn}
\usepackage{amsmath,amssymb,amsfonts}
\usepackage{textcomp}
\usepackage{xspace}
\usepackage{comment}
\usepackage[printonlyused,nohyperlinks]{acronym}
\usepackage{xurl}
\usepackage{bookmark}
\providecommand{\burl}[1]{\url{#1}}
\newcommand{\Lc}{\Lambda_c^+}

\newcommand{\LcLc}{\Lambda_c^+\bar{\Lambda}_c^-}

\newcommand{\ParT}{Particle Transformer}

\begin{document}

\title[Transferable event representations at BESIII]{Learning transferable event representations for charmed baryon physics at BESIII}

\author[1]{\fnm{Kaixuan} \sur{Huang}}\email{huangkaixuan@ucas.ac.cn}
\author[1]{\fnm{Yangu} \sur{Li}}\email{liyangu@ucas.ac.cn}
\author[1]{\fnm{Junpeng} \sur{Zhao}}\email{zhaojunpeng19@mails.ucas.ac.cn}
\author*[1]{\fnm{Peilian} \sur{Li}}\email{lipeilian@ucas.ac.cn}
\author[2]{\fnm{Peirong} \sur{Li}}\email{prli@lzu.edu.cn}
\author*[1]{\fnm{Xiao-Rui} \sur{Lyu}}\email{xiaorui@ucas.ac.cn}
\author*[1]{\fnm{Yunxuan} \sur{Song}}\email{songyunxuan@ucas.ac.cn}
\author[3,1]{\fnm{Shengsen} \sur{Sun}}\email{sunss@ihep.ac.cn}
\author[1]{\fnm{Yangheng} \sur{Zheng}}\email{zhengyh@ucas.ac.cn}

\affil[1]{\orgname{University of Chinese Academy of Sciences}, \orgaddress{\city{Beijing}, \postcode{100049}, \country{China}}}
\affil[2]{\orgname{Lanzhou University}, \orgaddress{\city{Lanzhou}, \postcode{730000}, \country{China}}}
\affil[3]{\orgname{Institute of High Energy Physics, Chinese Academy of Sciences}, \orgaddress{\city{Beijing}, \postcode{100049}, \country{China}}}

\abstract{
Deep learning has become an essential tool in high-energy physics, where the ability to learn transferable event representations can significantly improve model generalization across related physics processes.
In this work, we present a Particle Transformer-based framework for learning such representations for charmed baryon physics in the BESIII experiment. 
The framework is implemented through large-scale pre-training on Monte Carlo simulation samples and subsequent fine-tuning for downstream analyses.
Using the production and decays of the charmed baryon $\Lc$ as a benchmark, we develop pre-trained models for both event classification and momentum-direction regression.
The classification model learns discriminative event representations for the dominant physics categories, rejecting $96.90\%$ of background events at a signal efficiency of 90.00\%.
Across 12 benchmark $\Lc$ decay channels, fine-tuning from the pre-trained model achieves performance comparable to or better than training from scratch, with particularly clear improvements in low-statistics regimes.
For the regression task, the pre-trained model improves the average angular resolutions across the same benchmark channels. Further improvement is obtained after fine-tuning in the representative semileptonic decay $\Lambda_c^+ \to p K^- e^+ \nu_e$. In addition, the predicted momentum direction enables reconstruction of the parent kinematics without a recoil-side tag and thereby increases signal efficiency. For this semileptonic decay, the pre-trained regression model increases the signal efficiency by approximately one order of magnitude relative to the double-tag method.
This strategy provides a scalable solution for a wide
range of physics cases at BESIII and can be extended to other high energy experiments.



}


\maketitle
\clearpage

\section{Introduction}
\label{sec:introduction}
Machine learning has become an indispensable tool in high-energy physics~(HEP), with applications spanning detector simulation, event reconstruction, and physics analysis~\cite{ML_in_HEP, Living_Review_ML,DL,DL_HEP, ML_nature, Roadmap, HEP_Software_Foundation}.
In particular, deep learning methods have demonstrated strong performance in event classification~\cite{Classification_1} and regression tasks, enabling significant improvements in signal--background discrimination and in the reconstruction of physical observables such as energy, momentum and vertex position~\cite{energy_regression, Vertex_energy_reconstruction}. 


In the BESIII experiment~\cite{BESIII_track_reconstruction,Lc2nenv}, deep learning models based on the \ParT{}~\cite{ParT, jet_tagging, weaver} architecture have achieved substantial gains in several analyses, including rare decay searches and precision measurements~\cite{D2GammaENv, Lc2petaPrime, Lc2ppi0}. These approaches have reduced background levels by more than two orders of magnitude, leading to enhanced signal purity and improved measurement sensitivity.

Despite these successes, current applications at BESIII predominantly rely on training models from scratch for individual physics analyses. This paradigm leads to duplicated development effort across decay channels with similar topologies and detector conditions, and it limits performance in low-statistics scenarios where training data are insufficient. 


A natural solution to these challenges is transfer learning through pre-training, in which a model is first trained on large and diverse datasets to learn general-purpose representations and is subsequently fine-tuned for specific downstream tasks. This paradigm has proven highly effective in fields such as natural language processing~\cite{gpt1, gpt5} and bioinformatics~\cite{ESM1,ESM2}, and recent studies relevant to HEP experiments such as CMS~\cite{CMS} and ATLAS~\cite{ATLAS} have demonstrated its potential for improving classification performance and data utilization efficiency~\cite{CMS_pre_training, ATLAS_pre_training}.

In this work, 
we propose a general transfer learning strategy for the BESIII experiment based on the \ParT{} architecture. 
Using $\Lc$ baryon physics~\cite{LcPhysics} as a benchmark, we construct pre-trained models for event classification and momentum-direction regression, and evaluate their performance across multiple decay channels through direct application and fine-tuning. 
This benchmark is physically well motivated: different $\Lambda_c^+$ decay channels are produced from the same $e^+e^- \to \Lambda_c^+\bar{\Lambda}_c^-$ source and share common event-level topological structures, such as back-to-back charmed-baryon production near threshold, multi-prong charged final states, neutral cluster patterns, 
and missing-kinematic constraints from the known center-of-mass energy.
These common structures provide a natural basis for learning transferable event representations through pre-training.

This strategy provides an extensible approach for deep learning applications at BESIII and beyond.
Throughout this paper, charge-conjugate processes are implied.



The remainder of this paper is organized as follows.
Section~\ref{sec:pretrain_besiii} introduces the BESIII experiment, related deep learning applications at BESIII, and developments in transfer learning through pre-training for HEP experiments.
Section~\ref{sec:Methodologies} presents the general transfer learning methodology, including the input representation of event information, the \ParT{}-based architecture, the design of the pre-training tasks for event classification and momentum-direction regression, and the fine-tuning strategy for downstream tasks.
Section~\ref{sec:application to Lc physics} describes the application of this approach to $\Lc$ physics.
Section~\ref{sec:Performance} discusses the systematic performance evaluation of the pre-trained models for both event classification and momentum-direction regression.
Finally, Section~\ref{sec:summary} summarizes the main results.

\section{The BESIII experiment and transfer learning}
\label{sec:pretrain_besiii}
\subsection{The BESIII experiment}
\label{subsec:besiii}
The BESIII detector~\cite{BESIII_Detector,bes_unity} records symmetric $e^+e^-$ collisions
provided by the BEPCII storage rings~\cite{BEPCII}
in the center-of-mass energy range from 1.84 to 4.95~GeV, with a peak luminosity of $1.1 \times 10^{33}\;\text{cm}^{-2}\text{s}^{-1}$
achieved at $\sqrt{s} = 3.773\;\text{GeV}$.
BESIII has collected large data samples in this energy region~\cite{BESIII_Future, DQM, OEC, minghua_bes_datasets}.
The cylindrical core of the BESIII detector covers 93\% of the full solid angle
and consists of a helium-based multilayer drift chamber~(MDC),
a time-of-flight system~(TOF),
and a CsI(Tl) electromagnetic calorimeter~(EMC).
These subdetectors provide measurements used for charged-track reconstruction,
charged-particle identification, and the reconstruction of EMC clusters.
They are enclosed in a superconducting solenoidal magnet
providing a 1.0~T magnetic field.
The solenoid is supported by an octagonal flux-return yoke
with resistive plate counter muon identification modules
interleaved with steel.


The physics program of the BESIII experiment covers a broad range
of topics in the $\tau$-charm energy region~\cite{bes_physics,BESIII_Future, Li:2025nzx},
including studies on the light hadrons,
charmonium and charmonium-like states, as well as charmed meson and baryon decays, measurements of $R$ values, QCD observables and $\tau$ mass,
and searches for new physics beyond the Standard Model~(SM).

\subsection{Deep learning and transfer learning }
\label{sec:dl_besiii}
Current deep learning applications at BESIII predominantly follow a training-from-scratch paradigm, 
in which dedicated models are developed for the decay channels of interest.
Compared with conventional cut-based methods, 
these models can significantly reduce the background level while maintaining signal efficiency, therefore improving signal purity~\cite{Lc2nenv, D2GammaENv, Lc2petaPrime, Lc2ppi0}.

Nonetheless, this paradigm has several limitations. 
Due to the huge parameter space in deep learning models, effective model training typically requires at least several tens of thousands of signal and background simulation events to achieve reliable performance. However, production of such sample sizes are challenging and not always available for many processes.
Moreover, similar decay channels often require separate model development from scratch, leading to redundant effort.
Especially for the study of $\Lc$ physics at BESIII, which involves tens of decay channels,
there are substantial duplications in task-specific development~\cite{Lc2nenv, Lc2petaPrime, Lc2ppi0}.


Pre-training has recently emerged as an effective transfer-learning strategy in HEP, 
targeting different physics tasks~\cite{CMS_pre_training, ATLAS_pre_training, Pre-training_NST,FM1,FM2}. 
In jet classification, contrastive pre-training on the large-scale JetClass dataset has been used to learn transferable jet representations and has shown improved performance over training from scratch after fine-tuning~\cite{CMS_pre_training}. 
In event classification, pre-training on large simulated samples of Higgs-boson and top-quark production processes has improved downstream event classification, especially in low-statistics regimes~\cite{ATLAS_pre_training}. 
These activities motivate the development of a general transfer learning strategy tailored to BESIII, where multiple physics channels share similar detector responses and event topologies.



\section{Methodologies}
\label{sec:Methodologies}
Transfer learning through pre-training offers significant advantages for learning transferable representations and improving model performance.
However, constructing a unified transfer learning strategy for BESIII entails several challenges.
One key challenge is the construction of a general event representation.
The input features should not rely on variables defined only for specific physics objects, decay channels, or signal definitions, 
while the model architecture must accommodate events with variable final-state particle multiplicities.
Another challenge is the adaptation of the pre-trained model to analysis-specific downstream tasks.
A suitable fine-tuning strategy is required to tailor the general representations learned during pre-training to each target physics analysis.
The transfer learning strategy at BESIII consists of three main components: event input representation, the \ParT{}-based model architecture, and fine-tuning.








\begin{figure*}[!htb]
	\centering
	\includegraphics[width=0.95\hsize]{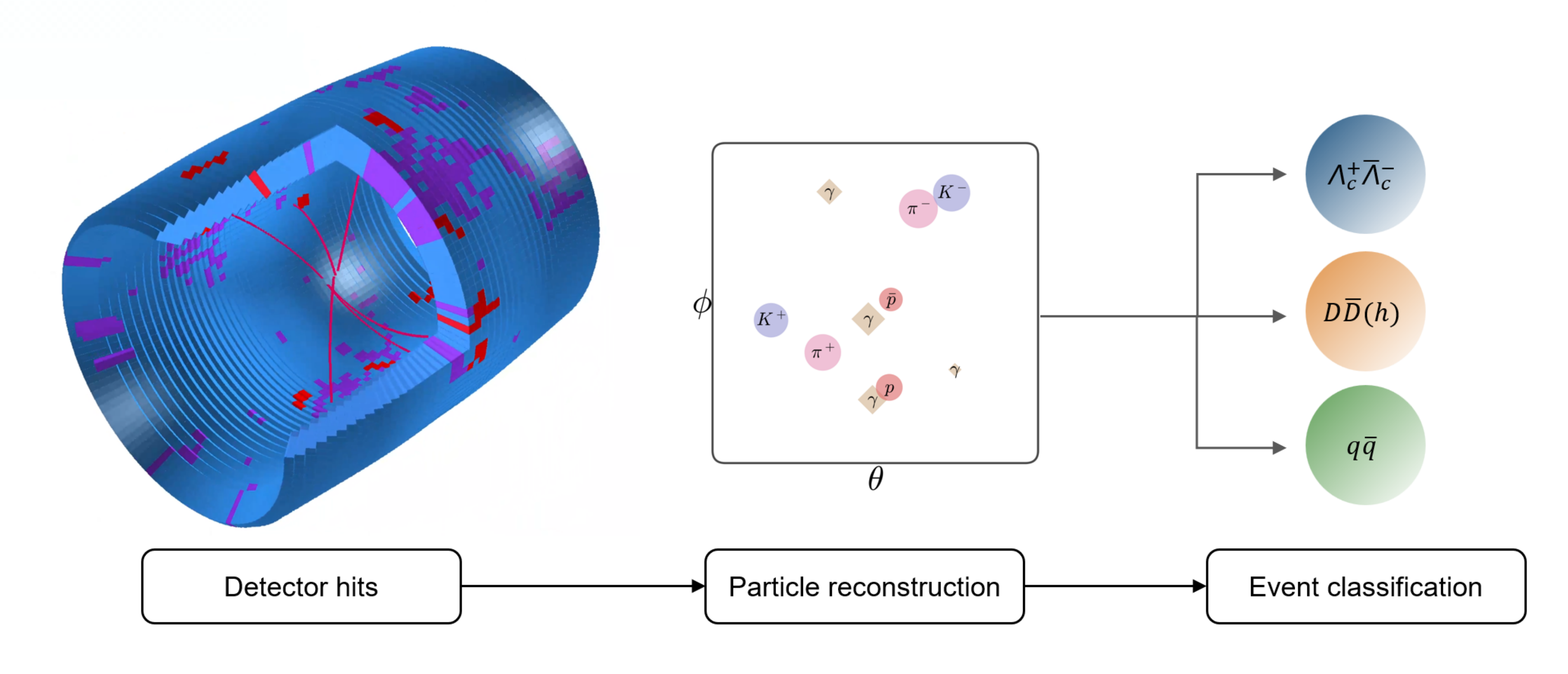}
	\caption{Schematic illustration of the event input representation at BESIII.}
	\label{fig:event_input}
\end{figure*}

\subsection{Event input representation}
\label{subsec:event input representation}

In HEP experiments, event information is typically organized into the hit level (raw detector responses such as timing, charge, and energy deposition), the object level (reconstructed physics objects such as charged tracks and calorimeter clusters, together with their kinematic properties), and the event level (global quantities such as total energy, missing energy, and particle multiplicities).

At BESIII, the input representation is constructed at the object level, 
since hit-level information is computationally prohibitive owing to the large number of detector channels,
whereas event-level observables are too compressed to preserve the topological structure of the final state.
As schematically shown in Fig.~\ref{fig:event_input},
raw detector hits are processed through particle reconstruction into charged tracks and EMC clusters,
which are organized as a point cloud and serve as input for pre-training models for tasks such as event classification.
Specifically, for each charged track, the input features comprise the azimuthal and polar angles, the charge, the momentum magnitude, and the parameters characterizing its helical trajectory in the MDC.
For each cluster, the input features comprise the azimuthal and polar angles, the energy deposition, the count of fired crystals, and the parameters describing its expansion scope among nearby crystals.
In addition, detector-response observables from the MDC, TOF, and EMC are incorporated as implicit information for particle identification and reconstruction quality.

These features correspond to general detector-response observables that are broadly available at BESIII and are not tied to any specific physics channel.
This generality allows the same set of input features to be transferred to a wide range of physics processes at BESIII.

\subsection{Model architecture}
\label{subsec:model architecture}
The \ParT{} architecture is adopted for this framework.
Its self-attention mechanism naturally handles unordered particle sets and is therefore well suited to point-cloud inputs. 
Moreover, it enables each particle to interact directly with all other particles in the event through attention weights, allowing the model to capture global correlations among particles.
The full model architecture consists of the \ParT{} backbone network and an output layer.
The backbone network comprises an embedding layer and an attention module.

In the embedding stage, 
charged track features and neutral cluster features are 
projected into a high-dimensional latent space.
In parallel,
pairwise interaction features are constructed for each particle pair and encoded into an interaction matrix $\mathbf{U_{BES}}$ through a separate embedding layer.
The interaction features are designed at BESIII to account for the spherical geometry of $e^+e^-$ collisions and energy-momentum conservation in three-dimensional space.
The interaction matrix $\mathbf{U_{BES}}$ is incorporated as a bias term in the attention weight computation of each particle attention block,
thereby enabling the model to directly exploit pairwise kinematic correlations between particles.
Specifically,
\begin{equation}
\mathbf{U_{BES}}=
\left\{
\begin{aligned}
&1-\cos\theta_{a,b},\\
&\min(E_a,E_b)\,(1-\cos\theta_{a,b}),\\
&\min(E_a,E_b)/(E_a+E_b),\\
&(E_a+E_b)^2-\left\|\vec{p}_a+\vec{p}_b\right\|^2,
\end{aligned}
\right.
\end{equation}
where $E_a$ and $E_b$ are the energies of the two particles,
$\vec{p}_a$ and $\vec{p}_b$ are their momentum vectors,
and $\theta_{a,b}$ is the opening angle between them in the center-of-mass frame.

The attention module consists of particle attention blocks followed by class attention blocks.
The particle attention blocks refine particle representations through multi-head self-attention,
enabling each particle to interact with all other particles in the event.
The class attention blocks employ a learnable class token that attends to all particle representations,
aggregating the variable-length particle-level information into a fixed-dimensional event-level representation.
The output layer maps the event-level representation to task-specific predictions through fully connected layers, with its configuration adapted to the target task, such as event classification or momentum-direction regression.





\subsection{Fine-tuning}
\label{subsec:fine-tuning}
Fine-tuning is a standard paradigm in transfer learning~\cite{Transfer_Learning}.
It initializes the model with the parameters of the pre-trained network
and further optimizes it on a downstream dataset,
so that it can be adapted to a specific physics analysis target.
By inheriting the representations learned during pre-training,
fine-tuning generally achieves higher data utilization efficiency and improved performance compared with training from scratch.
This property is particularly beneficial for decay channels with limited statistics.

A differential learning-rate strategy is adopted during fine-tuning.
Specifically,
a smaller learning rate is applied to the backbone network
to preserve the representations learned during pre-training as much as possible,
whereas a larger learning rate is used for the output layer
so that it can adapt more rapidly to the downstream objective.

\section{Application to \texorpdfstring{$\Lambda_c^+$}{Lambda-c+} physics}
\label{sec:application to Lc physics}

\subsection{Physics motivation}
\label{sec:physics motivation}
As the lightest charmed baryon, with a mass of $2286.46~\mathrm{MeV}/c^2$~\cite{pdg:2025},
the $\Lambda_c^+$ consists of a charm quark and two light quarks~($udc$).
It plays a fundamental role in charmed baryon spectroscopy and weak decay studies.

At BESIII,
$\Lambda_c^+$ physics offers distinct experimental advantages, owing to the large $e^+e^-$ data samples accumulated near the $\Lambda_c^+\bar{\Lambda}_c^-$ production threshold, where $\Lambda_c^+\bar{\Lambda}_c^-$ pairs are produced without additional hadrons,
providing a comparatively clean event environment
and favorable conditions for the application of single-tag (ST) and double-tag (DT) techniques~\cite{double_tag}.

Moreover, with nearly one hundred decay modes listed in the PDG~\cite{pdg:2025},
$\Lambda_c^+\bar{\Lambda}_c^-$ physics provides a suitable application scenario for transfer learning to reduce repeated model development across channels.
At BESIII, these decay channels share comparable detector responses and event topologies, 
making $\Lambda_c^+$ physics an ideal benchmark for validating the generic transfer learning strategy.

\subsection{Datasets}
\label{sec:datasets}
In this work,
the dataset is constructed from Monte Carlo~(MC) simulation samples~\cite{GEANT4, evtgen1, evtgen2, lundcharm1, lundcharm2}
covering multiple center-of-mass energy points in the range of $\sqrt{s} = 4.600$--$4.700\;\text{GeV}$.
The MC simulation configuration is consistent with that used in published BESIII physics analyses~\cite{Lc2nenv, D2GammaENv, Lc2ppi0, Lc2petaPrime, BOSS}.
The inclusive sample is partitioned by source file into training and validation sets for pre-training, training and validation sets for downstream tasks, and an independent test set, with respective fractions of 45\%, 5\%, 30\%, 10\%, and 10\%.

\textit{Pre-training datasets.}
For the event classification pre-training model,
the dataset is constructed to distinguish among the dominant physics categories
relevant to $\Lambda_c^+$ studies.
Events are divided into three categories:
the $\Lambda_c^+\bar{\Lambda}_c^-$ category,
the $D\bar{D}(h)$ category,
and the $q\bar{q}$ category.
The numbers of events in the three categories are balanced,
with a total of approximately $4.1\times10^7$ events.
The $\Lambda_c^+\bar{\Lambda}_c^-$ category consists of events produced via $e^+e^-\to\Lambda_c^+\bar{\Lambda}_c^-$.
The $D\bar{D}(h)$ category comprises events from the inclusive charm production in this energy region,
where $D$ denotes $D^0$, $D^+$, or $D_s^+$, and $h$ denotes one or more light hadrons, such as $\pi$ or $K$.
This category may also support future studies of $D$-related processes.
The $q\bar{q}$ category comprises hadronic background events excluding those assigned to the $D\bar{D}(h)$ category.
For the momentum-direction regression pre-training model,
only $\Lambda_c^+\bar{\Lambda}_c^-$ events from the classification pre-training sample are used,
corresponding to a total of approximately $1.4\times10^7$ events.

\textit{Downstream datasets.}
The datasets for downstream tasks depend on the specific physics channel.
To systematically evaluate the transferability and generalization performance of the pre-trained models across different decay channels,
twelve copious $\Lambda_c^+$ hadronic decays~\cite{12tag} are adopted as a benchmark set.
These channels are commonly used as tag modes in DT analyses:
$\Lambda_c^+ \to pK_S^0$,
$pK^-\pi^+$,
$pK_S^0\pi^0$,
$pK_S^0\pi^+\pi^-$,
$pK^-\pi^+\pi^0$,
$\Lambda\pi^+$,
$\Lambda\pi^+\pi^0$,
$\Lambda\pi^+\pi^-\pi^+$,
$\Sigma^0\pi^+$,
$\Sigma^+\pi^0$,
$\Sigma^+\pi^+\pi^-$,
and $p\pi^+\pi^-$.
For each benchmark channel,
channel-specific selections from the downstream and independent test partitions are used to evaluate three training strategies:
direct application of the pre-trained model,
fine-tuning of the pre-trained model,
and training from scratch.
In downstream event classification,
events are reclassified according to the signal process of the specific physics analysis,
with the dataset for each benchmark channel grouped into three event categories:
signal events,
defined as $\Lambda_c^+\bar{\Lambda}_c^-$ events in which at least one of the two baryons decays through the target decay channel;
$\Lambda_c^+\bar{\Lambda}_c^-$ background events,
consisting of $\Lambda_c^+\bar{\Lambda}_c^-$ events excluding the target decay channel;
and hadronic background events,
comprising both $D\bar{D}(h)$ and $q\bar{q}$ events.
The classification score is the $\Lambda_c^+\bar{\Lambda}_c^-$ class probability for the directly applied pre-trained model and the target-signal probability for the fine-tuned and from-scratch models.
The training sample sizes for different benchmark channels range approximately from $2\times10^4$ to $2.8\times10^6$,
depending on the number of samples retained after processing by the analysis software package for each decay channel.
In the direct application of the pre-trained momentum-direction regression model,
the same 12 benchmark channels are used to evaluate the generality of the pre-trained model across different decay topologies.

In addition,
to examine the applicability of the momentum-direction regression model to final states containing undetected neutral particles,
the semileptonic decay $\Lambda_c^+ \to pK^-e^+\nu_e$~\cite{Lc2pKenv} is considered as a representative case.
This dataset contains approximately $8.5\times10^4$ events and is evaluated under all three training strategies.





\subsection{Experimental setup}
\label{sec:experimental setup}
The pre-training model adopts a deeper network configuration than that used in previous analyses~\cite{Lc2nenv, D2GammaENv, Lc2petaPrime, Lc2ppi0} to increase the model capacity.
Specifically,
the pre-training model employs 8 attention heads,
8 particle-attention layers,
and 3 class-attention layers,
compared with 4 particle-attention layers and 2 class-attention layers in previous studies.
This configuration enables the model to learn more expressive representations from the pre-training dataset.

In the embedding layers,
particle features are encoded by a three-layer perceptron with 96, 384, and 96 nodes per layer,
whereas pairwise features are encoded by four pointwise one-dimensional convolutional layers with 64, 64, 64, and 8 channels.
To mitigate overfitting,
a dropout rate of 0.1 is applied.

Classification and regression are trained as separate models. The classification model outputs three probabilities and is optimized using cross-entropy loss. The regression model directly predicts the four laboratory-frame angles of the $\Lambda_c^+$ and $\bar{\Lambda}_c^-$ baryons and is optimized using a log-cosh loss averaged over the batch.
All models are optimized using the Ranger optimizer~\cite{Ranger}.

Both pre-training models are trained for 50 epochs with a batch size of 8192.
The initial learning rates are $5\times10^{-6}$ and $1\times10^{-5}$, and the peak learning rates are $5\times10^{-4}$ and $1\times10^{-3}$ for classification and regression, respectively.
The learning-rate schedule uses linear warm-up, a constant-rate phase, and cosine decay.
Training is performed on NVIDIA L20 GPUs.

During fine-tuning,
the weights of the best pre-trained model are loaded as initialization,
and a learning-rate multiplier of 10 is applied to the output layer.
This implements the differential learning-rate strategy described in Section~\ref{subsec:fine-tuning}.
For each downstream task, fine-tuning and training from scratch use the same network architecture and data partitions and are performed for 50 epochs with batch size 1024.
Both strategies use an initial learning rate of zero and the same learning-rate schedule as pre-training.
The peak learning rates are $1\times10^{-4}$ for the fine-tuned backbone and $1\times10^{-3}$ for its output layer, and $1\times10^{-3}$ for both the backbone and output layer when training from scratch.

\begin{figure*}[!htb]
  \centering
\includegraphics[width=0.48\textwidth]{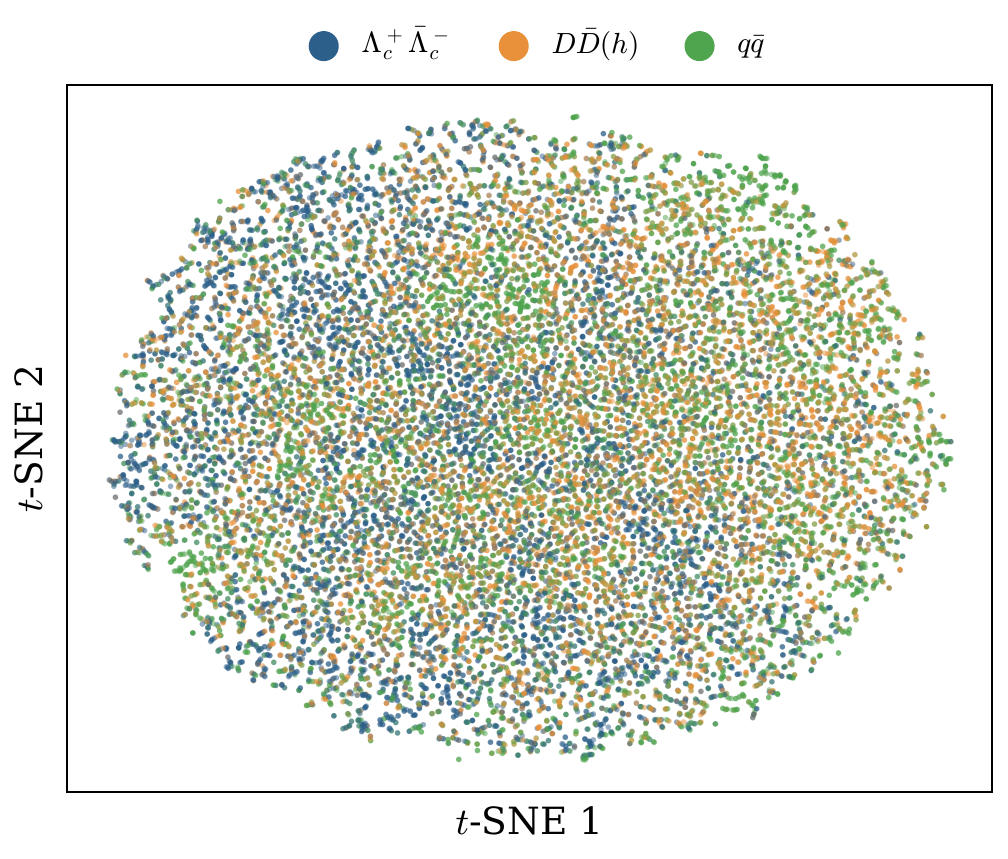}
\includegraphics[width=0.48\textwidth]{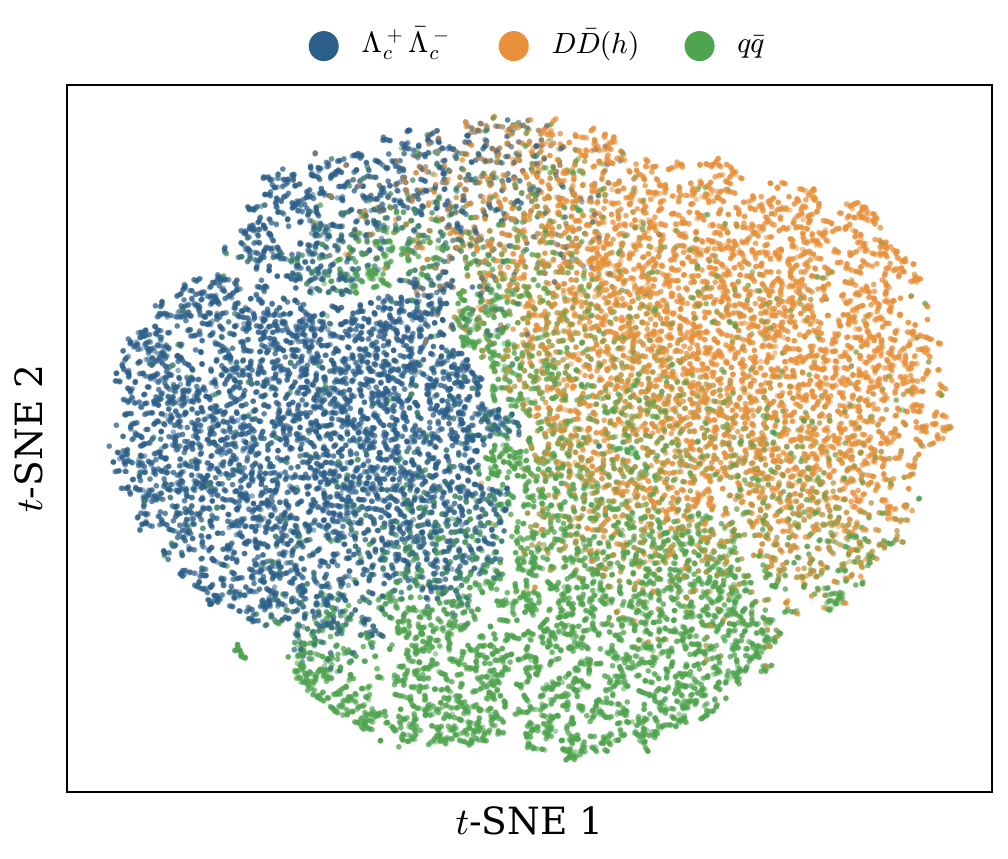}
  \caption{
Visualization of the learned event representations using $t$-SNE.
Left: the model at initialization (before training).
Right: the model after pre-training (best epoch).
The representations are extracted from the normalization layer
after the last attention layer in the backbone network,
evaluated on the inclusive dataset.
}
\label{fig:tSNE}
\end{figure*}

\section{Performance evaluation}
\label{sec:Performance}
The two pre-trained models for event classification and momentum-direction
regression are systematically evaluated, and the corresponding fine-tuned models
are compared with the models trained from scratch.
Within each task, all methods are compared using the same independent test sample. For each task, pre-training results are reported for the model with the best validation performance among three runs with different random seeds. Fine-tuned and from-scratch results are obtained from the best model among five runs with different random seeds.

\subsection{Event classification}
\label{sec:event classification}

The pre-trained classification model demonstrates strong discriminative performance on the inclusive dataset. 
As shown in Fig.~\ref{fig:tSNE}, the visualization of learned representations, a two-dimensional projection of the normalization layer's outputs obtained via $t$-distributed stochastic neighbor embedding ($t$-SNE)~\cite{tSNE, tSNE_ref1, tSNE_ref2}, reveals clear separation among different physics categories after pre-training.
This separation illustrates the category structure learned during supervised pre-training.

\begin{figure}[!htb]
	\centering
	\includegraphics[width=0.95\linewidth]{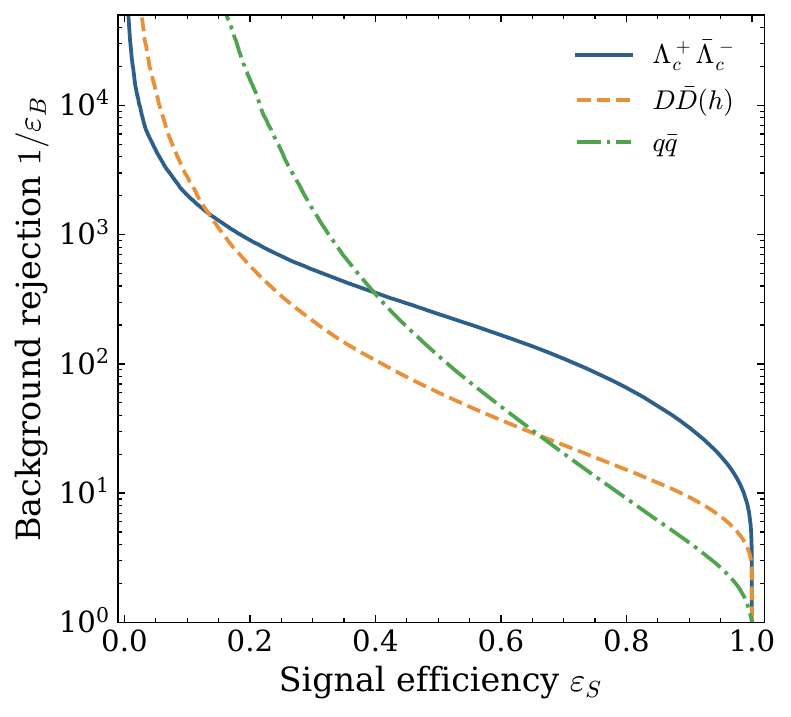}
	\caption{
  ROC curves of the pre-trained event classification model for the $\LcLc$, $D\bar{D}(h)$, and $q\bar{q}$ categories, evaluated on the independent, class-balanced inclusive test sample in a one-versus-rest manner.
  }
	\label{fig:pre-training-event classification}
\end{figure}

The classification performance of the pre-trained model is further quantified
using receiver operating characteristic~(ROC) curves.
Figure~\ref{fig:pre-training-event classification} presents the ROC curves
for the $\LcLc$, $D\bar{D}(h)$, and $q\bar{q}$ categories evaluated on the independent, class-balanced inclusive test sample,
where the background rejection factor, defined as the reciprocal of the false positive rate~(FPR), $1/\mathrm{FPR}$, is plotted against the signal efficiency.
The corresponding rejection rate, $1-\mathrm{FPR}$, expresses the fraction of background events rejected at a given working point and is used in the discussion below.
The area under the ROC curve~(AUC) provides an overall measure of the discrimination power.

For the $\LcLc$ category,
the AUC reaches $0.988$.
At a signal efficiency of 90.00\%,
the model achieves a background rejection rate of $96.90\%$.
This result provides a solid foundation for the direct application of the pre-trained model in downstream analyses.
For the $D\bar{D}(h)$ category,
the AUC is $0.962$.
At a signal efficiency of 90.00\%,
the model achieves a background rejection rate of $89.27\%$.
This result indicates that the pre-trained model can also provide an effective basis for event selection in $D\bar{D}(h)$ physics studies in this energy region.
In addition, the AUC for the $q\bar{q}$ background category is $0.928$,
indicating good discrimination capability against the other event categories.

\begin{figure*}[!htb]
  \centering
\includegraphics[width=0.48\textwidth]{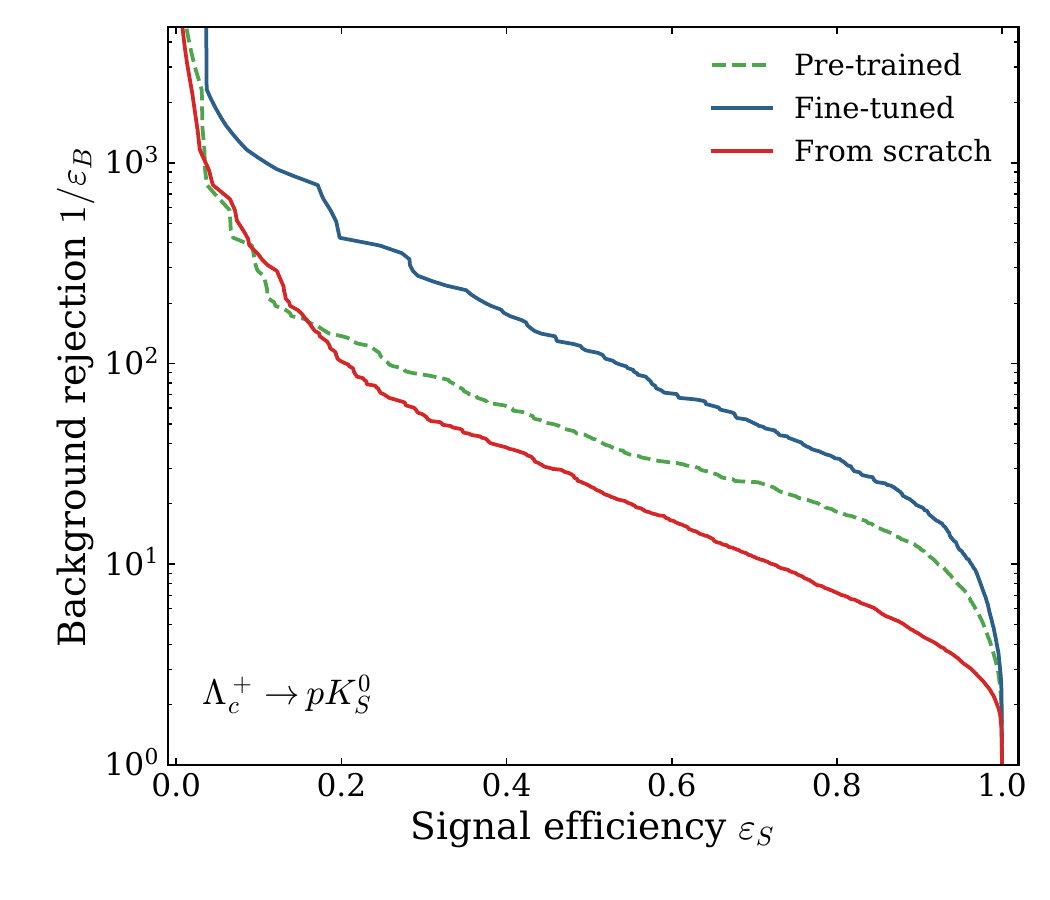}
\includegraphics[width=0.48\textwidth]{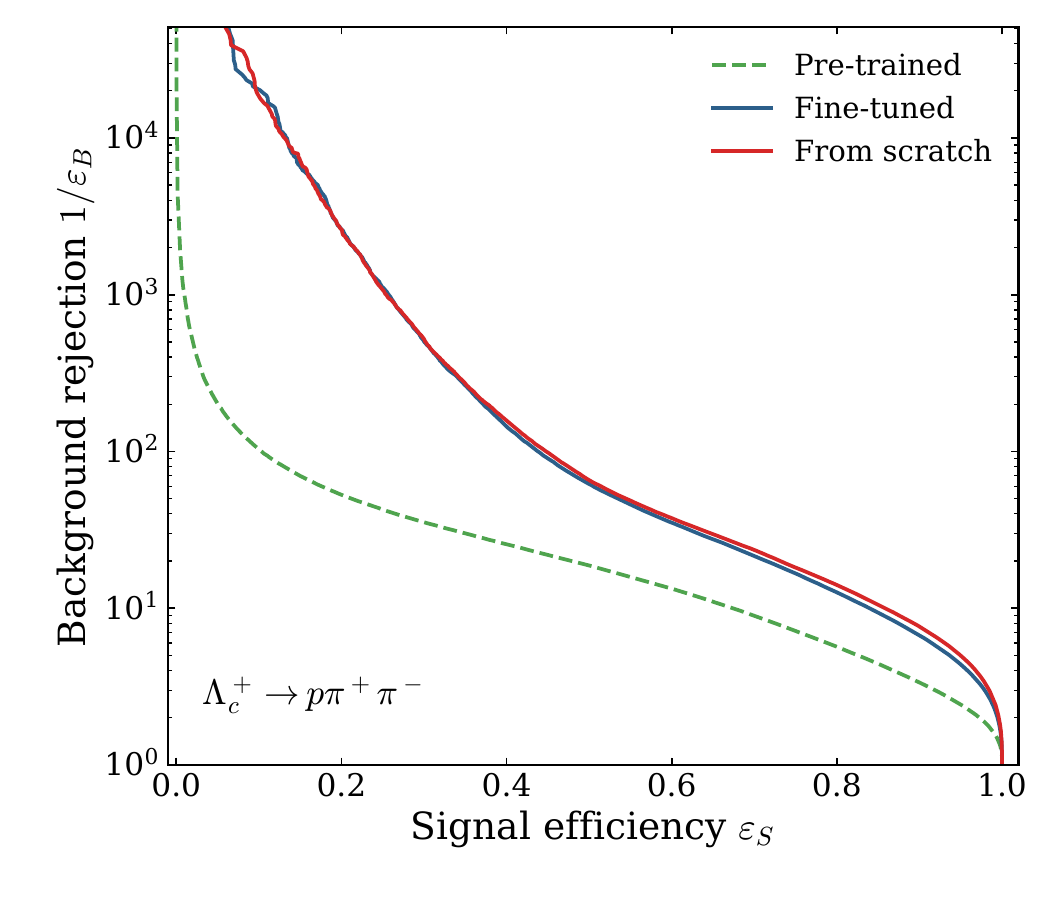}
  \caption{ROC curves for two benchmark channels with contrasting training sample sizes. The left and right panels show the results for $\Lc \to pK_S^0$ with $2.8 \times 10^4$ events and $\Lc \to p\pi^+\pi^-$ with $2.8 \times 10^6$ events, respectively.}

\label{fig:12tag-roc-comparison}
\end{figure*}

To further assess and compare the three training strategies, namely direct application of the pre-trained model, fine-tuning from pre-trained weights, and training from scratch,
systematic evaluations are performed across the 12 benchmark channels.
Figure~\ref{fig:12tag-roc-comparison} presents the ROC curves
for two representative benchmark channels with contrasting training sample sizes.
For the $\Lc \to pK_S^0$ channel,
with approximately $2.8 \times 10^4$ available training events,
the fine-tuned model achieves clearly superior signal--background discrimination
compared with the model trained from scratch,
demonstrating the effectiveness of the adopted fine-tuning strategy
in the low-statistics regime.
In contrast,
for the $\Lc \to p\pi^+\pi^-$ channel,
with approximately $2.8 \times 10^6$ available training events,
the fine-tuned model achieves comparable performance to the model trained from scratch.

\begin{figure*}[!htb]
	\centering
	\includegraphics[width=0.95\hsize]{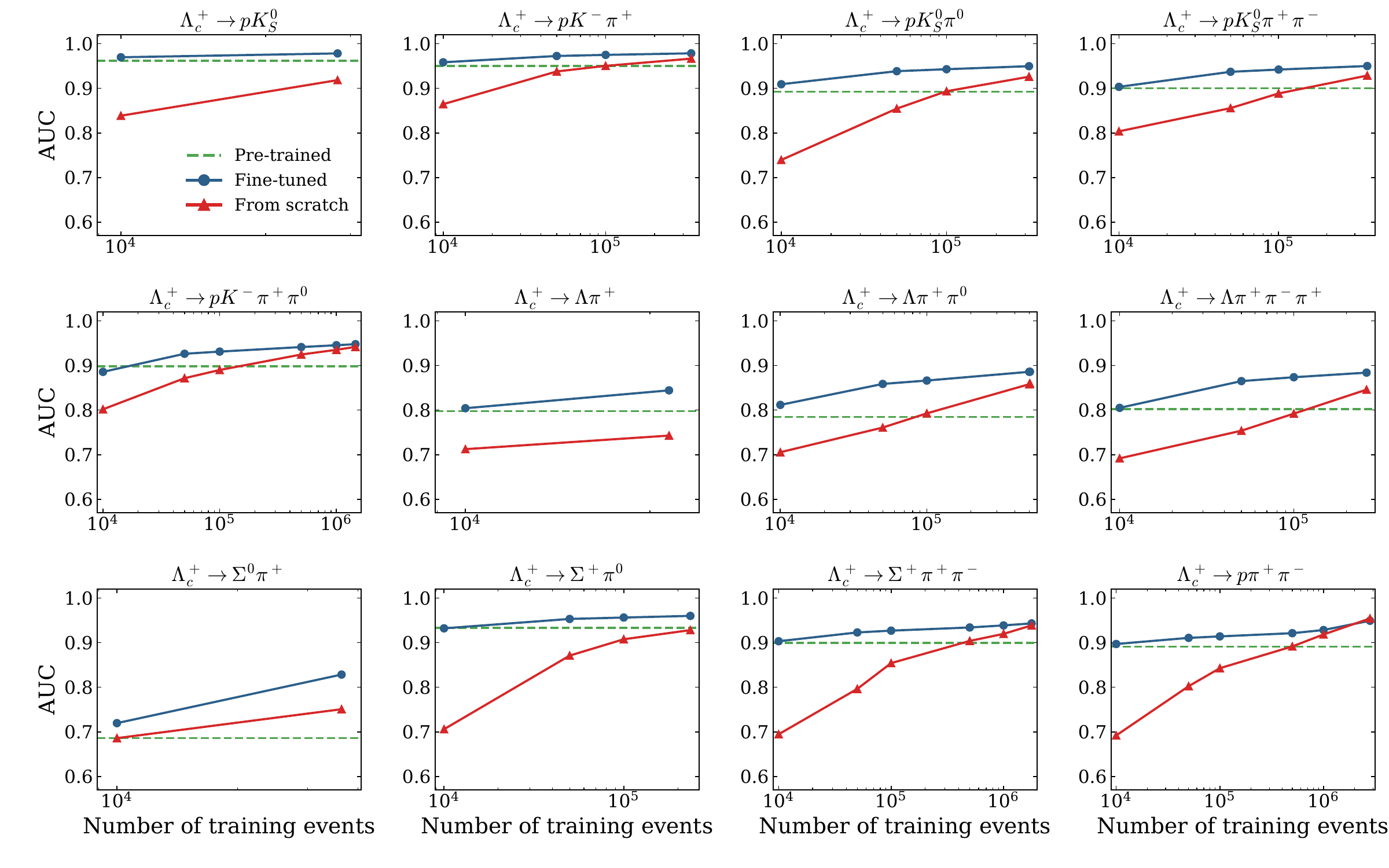}
	\caption{Comparison of AUC values with different sample sizes across the 12 benchmark channels for three training strategies.}
	\label{fig:auc-comparison-AUC-values}
\end{figure*}

To quantify this trend systematically across all 12 benchmark channels,
the AUC values obtained with the three training strategies are compared
as a function of training sample size
in Fig.~\ref{fig:auc-comparison-AUC-values}.
For each channel, the displayed training sample size is varied from $1 \times 10^4$ up to $2.8 \times 10^6$,
bounded by the total available samples of that channel.
The directly applied pre-trained model demonstrates good signal--background discrimination
for all 12 channels,
with AUC values ranging from 0.686 to 0.961.
Fine-tuning yields performance comparable to or better than that obtained by training from scratch,
with the improvement particularly pronounced for channels with
limited statistics, confirming the advantage of transfer learning in data-scarce
scenarios.

\begin{table*}[!htb]
\centering
\setlength{\tabcolsep}{3pt}
\caption{Comparison of the background rejection rate ($1-\mathrm{FPR}$, in \%) at 90.00\% signal efficiency for different training strategies across the 12 benchmark channels.}
\label{tab:bkg-eff-comparison}
\begin{tabular}{lccc}
\hline\hline
Decay channel & Pre-trained & Fine-tuned & From scratch \\
\hline
$\Lambda_c^+ \to pK_S^0$                  & $91.11$ & $94.70$ & $78.08$ \\
$\Lambda_c^+ \to pK^-\pi^+$             & $88.59$ & $93.78$ & $90.70$ \\
$\Lambda_c^+ \to pK_S^0\pi^0$            & $76.68$ & $85.44$ & $80.28$ \\
$\Lambda_c^+ \to pK_S^0\pi^+\pi^-$       & $76.08$ & $84.78$ & $79.20$ \\
$\Lambda_c^+ \to pK^-\pi^+\pi^0$         & $75.60$ & $83.41$ & $81.74$ \\
$\Lambda_c^+ \to \Lambda\pi^+$          & $50.94$ & $54.64$ & $37.31$ \\
$\Lambda_c^+ \to \Lambda\pi^+\pi^0$     & $52.50$ & $67.49$ & $60.75$ \\
$\Lambda_c^+ \to \Lambda\pi^+\pi^-\pi^+$& $53.91$ & $67.35$ & $57.47$ \\
$\Lambda_c^+ \to \Sigma^0\pi^+$         & $34.65$ & $52.49$ & $39.90$ \\
$\Lambda_c^+ \to \Sigma^+\pi^0$         & $80.61$ & $88.38$ & $78.17$ \\
$\Lambda_c^+ \to \Sigma^+\pi^+\pi^-$    & $72.04$ & $82.52$ & $81.07$ \\
$\Lambda_c^+ \to p\pi^+\pi^-$           & $70.26$ & $85.14$ & $86.96$ \\
\hline\hline
\end{tabular}
\end{table*}

This is further illustrated by the background rejection rates at a signal efficiency of 90.00\%,
which are summarized in Table~\ref{tab:bkg-eff-comparison} for all 12 benchmark channels.
The directly applied pre-trained model achieves rejection rates of 34.65--91.11\%, while fine-tuning improves these to 52.49--94.70\%, compared with 37.31--90.70\% for the model trained from scratch.
The fine-tuned model achieves the highest rejection rate in 11 of the 12 channels.

\subsection{Momentum-direction regression}
\label{sec:momentum direction regression}
For the momentum-direction regression model,
as described in Section~\ref{sec:datasets},
the regression targets are the polar angles ($\theta$) and azimuthal angles ($\phi$) of the $\Lambda_c^+$ and $\bar{\Lambda}_c^-$ baryons in the laboratory frame.
The model performance is evaluated using the residual distributions between the predicted values ($\hat{\theta}$, $\hat{\phi}$) and the corresponding MC truth ($\theta_{\mathrm{MC}}$, $\phi_{\mathrm{MC}}$).
The residuals are defined as
$\Delta\theta = \hat{\theta} - \theta_{\mathrm{MC}}$
and
$\Delta\phi = \hat{\phi} - \phi_{\mathrm{MC}}$, with $\Delta\phi$ wrapped to $[-\pi,\pi)$.
The mean $\mu$ and standard deviation $\sigma$ of the residual distributions are used to quantify the central bias and the resolution,
respectively.

\begin{figure*}[!htb]
  \centering
\includegraphics[width=0.48\textwidth]{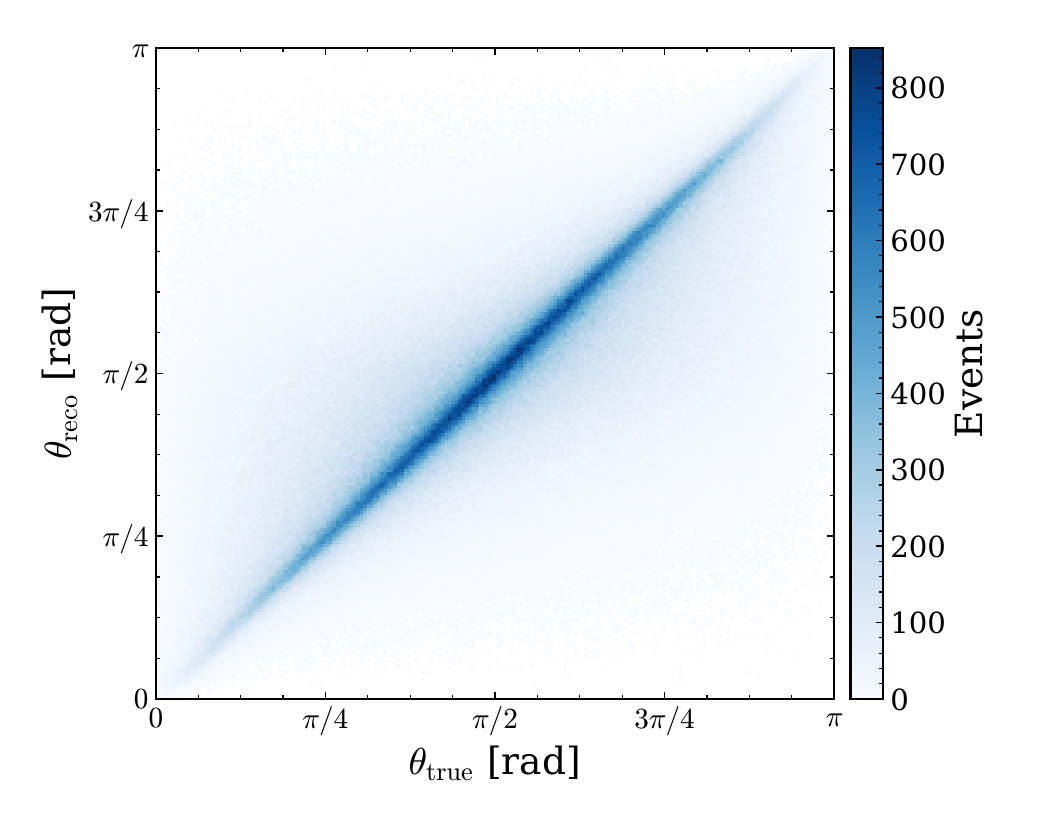}
\includegraphics[width=0.48\textwidth]{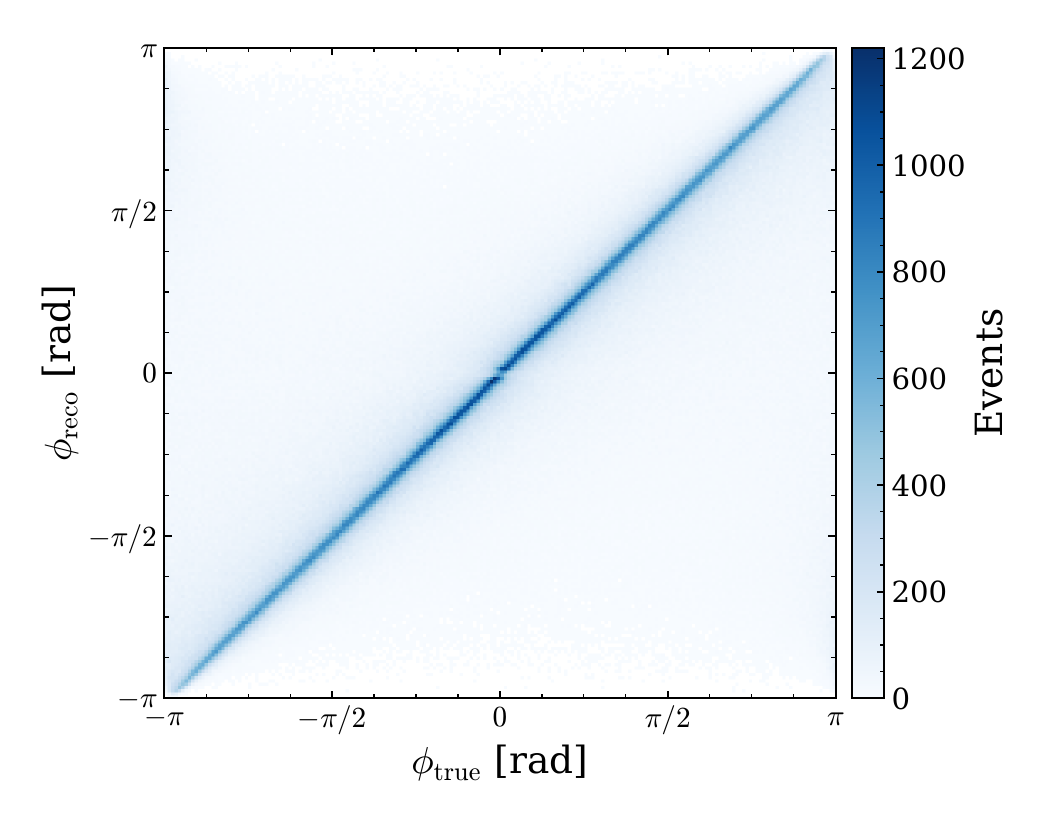}
\caption{Correlation between predicted and MC truth values of the momentum-direction angles for the best pre-trained regression model evaluated on the inclusive $\Lambda_c^+$ test sample. The left and right panels show $\theta$ and $\phi$, respectively.}
\label{fig:regression}
\end{figure*}

For the inclusive test sample shown in Fig.~\ref{fig:regression}, the angular residuals have mean values close to zero, with $\mu(\Delta\theta)=-0.001~\mathrm{rad}$ and $\mu(\Delta\phi)=0.000~\mathrm{rad}$.
The resolutions are
$\sigma(\Delta\theta)=0.425~\mathrm{rad}$
and
$\sigma(\Delta\phi)=0.896~\mathrm{rad}$,
demonstrating effective momentum-direction prediction capability.

\begin{figure*}[!htb]
	\centering
	\includegraphics[width=0.95\hsize]{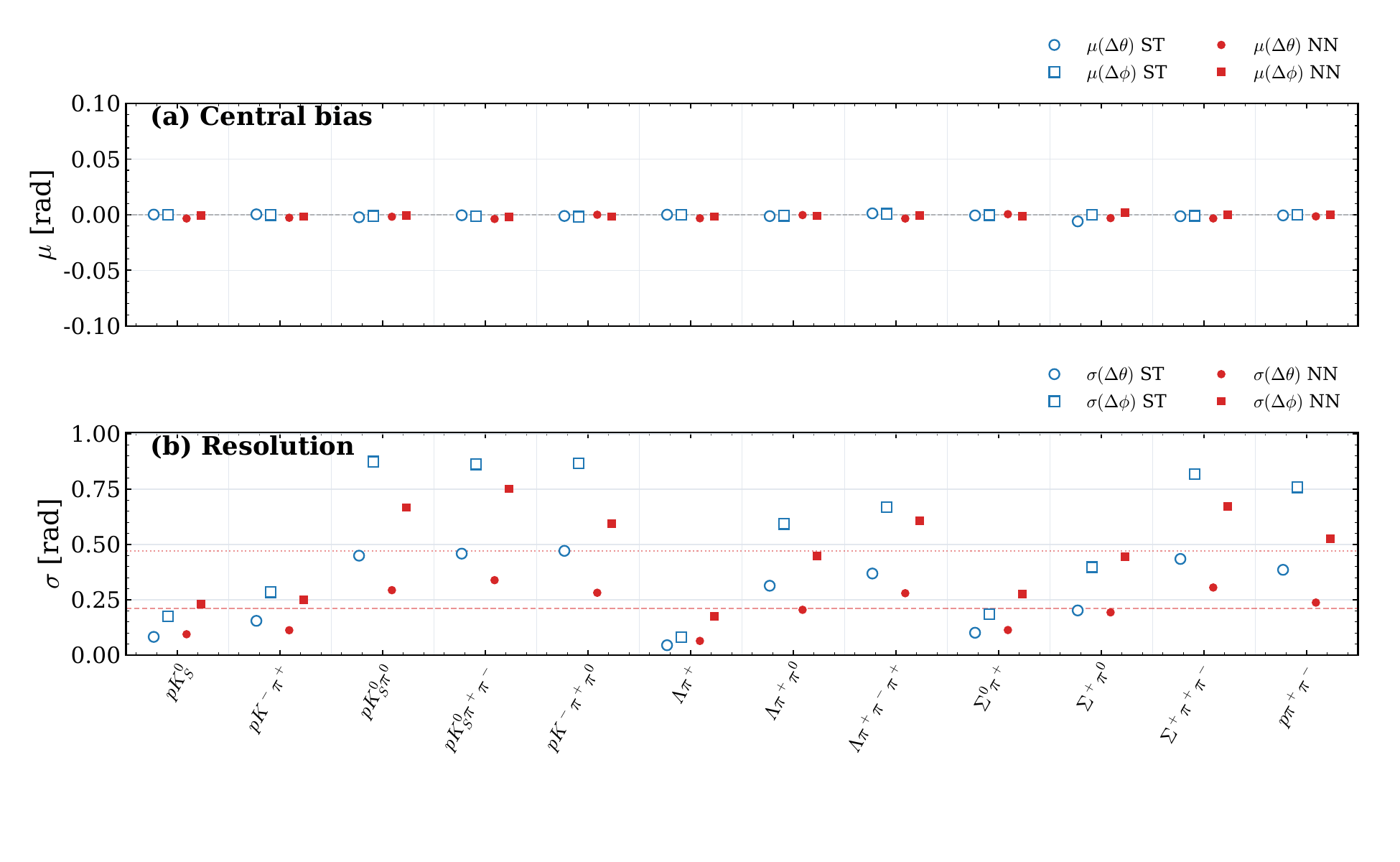}
	\caption{Central bias ($\mu$) and resolution ($\sigma$) of the pre-trained momentum-direction regression model across the 12 benchmark channels. The upper and lower panels show $\mu$ and $\sigma$ of the residual distributions, respectively. Red filled markers show pre-trained neural network (NN) results; blue open markers show ST reconstruction. 
  In the lower panel, the red dashed and dotted lines represent the average NN resolutions over the 12 channels for the polar ($\Delta\theta$) and azimuthal ($\Delta\phi$) angles, respectively.
  }
	\label{fig:regression_12tag_summary}
\end{figure*}

To further assess the generalization of the pre-trained regression model across different decay topologies,
the central bias and resolution are evaluated across the same 12 benchmark channels used in the classification task,
as shown in Fig.~\ref{fig:regression_12tag_summary}.
For each channel, the direction of the tagged charmed baryon is evaluated.
For all channels,
the central bias is close to zero, while the average resolutions over the 12 channels are
$0.210~\mathrm{rad}$ for $\Delta\theta$ and $0.470~\mathrm{rad}$ for $\Delta\phi$,
demonstrating stable performance across different decay topologies.

For comparison, 
the conventional ST reconstruction also shows
negligible central bias and gives average resolutions of
$\sigma(\Delta\theta)=0.289~\mathrm{rad}$ and
$\sigma(\Delta\phi)=0.547~\mathrm{rad}$
for the same 12 benchmark channels.
The regression model reduces the average $\Delta\theta$ and
$\Delta\phi$ resolutions by approximately $27.34\%$ and $14.11\%$, respectively,
relative to the conventional ST reconstruction.

One important application of the momentum-direction regression model is in physics processes with undetected particles in the final states,
which are challenging for conventional reconstruction methods.
Taking the semileptonic decay $\Lambda_c^+ \to p K^- e^+ \nu_e$ as an example,
the missing neutrino prevents the signal-side $\Lambda_c^+$ momentum direction
from being directly reconstructed.
In the conventional tag method, obtaining information on the $\Lambda_c^+$ momentum direction requires a fully reconstructed $\bar{\Lambda}_c^-$ decay,
which would largely compromise the limited signal statistics.
Taking $\Lambda_c^+ \to p K^- e^+ \nu_e$ as a benchmark channel,
we systematically compare the performance of the pre-trained model,
the fine-tuned model and the model trained from scratch.

The mean $\mu$ and standard deviation $\sigma$ of the residual distributions of $\Delta\theta$ and $\Delta\phi$ derived from these models are summarized in Table~\ref{tab:regression-semileptonic}.
The pre-trained model effectively predicts the momentum direction,
while maintaining negligible central bias.
The fine-tuned model further improves the resolutions,
with
$\sigma(\Delta\theta)=0.346~\mathrm{rad}$
and
$\sigma(\Delta\phi)=0.707~\mathrm{rad}$,
compared with
$\sigma(\Delta\theta)=0.483~\mathrm{rad}$
and
$\sigma(\Delta\phi)=1.035~\mathrm{rad}$
for the model trained from scratch.
The corresponding relative reductions in the $\Delta\theta$ and $\Delta\phi$
resolutions are $28.39\%$ and $31.69\%$, respectively.


\begin{table}[!htb]
\centering
\setlength{\tabcolsep}{2pt}
\caption{Comparison of the central bias and resolution of the momentum-direction regression for the decay $\Lambda_c^+ \to p K^- e^+ \nu_e$ under different training strategies. Values are given in units of rad.}
\label{tab:regression-semileptonic}
\begin{tabular}{l D{.}{.}{3} D{.}{.}{3} D{.}{.}{3}}
\hline\hline
Metric & \multicolumn{1}{c}{Pre-trained} & \multicolumn{1}{c}{Fine-tuned} & \multicolumn{1}{c}{From scratch} \\
\hline
$\mu(\Delta\theta)$ & -0.005 & 0.003 & 0.048 \\
$\sigma(\Delta\theta)$ & 0.380 & 0.346 & 0.483 \\
$\mu(\Delta\phi)$ & 0.002 & -0.005 & -0.009 \\
$\sigma(\Delta\phi)$ & 0.810 & 0.707 & 1.035 \\
\hline\hline
\end{tabular}
\end{table}


In addition, the regression model improves the signal efficiency for decays with undetected particles in the final state. Combining the predicted momentum direction with the beam energy constraint and the nominal parent mass yields the parent four-momentum without a recoil-side tag.

For the $\Lambda_c^+\to pK^-e^+\nu_e$ decay, the missing mass squared is calculated in the initial $e^+e^-$ center-of-mass frame using natural units ($c=1$) as
\begin{equation}
\begin{aligned}
M_{\rm miss}^{2}
 &=E_{\rm miss}^{2}-|\vec p_{\rm miss}|^{2}\\
 &=\left[E_{\rm beam}-(E_p+E_K+E_e)\right]^2\\
 &\quad-\left|\vec p_{\Lambda_c}^{\,\mathrm{reg}}
       -(\vec p_p+\vec p_K+\vec p_e)\right|^2,
\end{aligned}
\label{eq:pkenu-missing-mass}
\end{equation}
where $E_{\rm miss}$ and $\vec p_{\rm miss}$ are the missing energy and momentum, respectively, and $E_{\rm beam}=\sqrt{s}/2$ is the beam energy. The energies and momenta of the proton, $K^-$, and $e^+$ are denoted by $E_i$ and $\vec p_i$ ($i=p,K,e$), respectively. $\vec p_{\Lambda_c}^{\,\mathrm{reg}}$ denotes the parent momentum reconstructed from the predicted direction. $M_{\rm miss}^{2}$ is expected to be close to zero for signal events with an undetected neutrino.

\begin{figure}[!htbp]
\centering
\includegraphics[width=0.96\columnwidth]{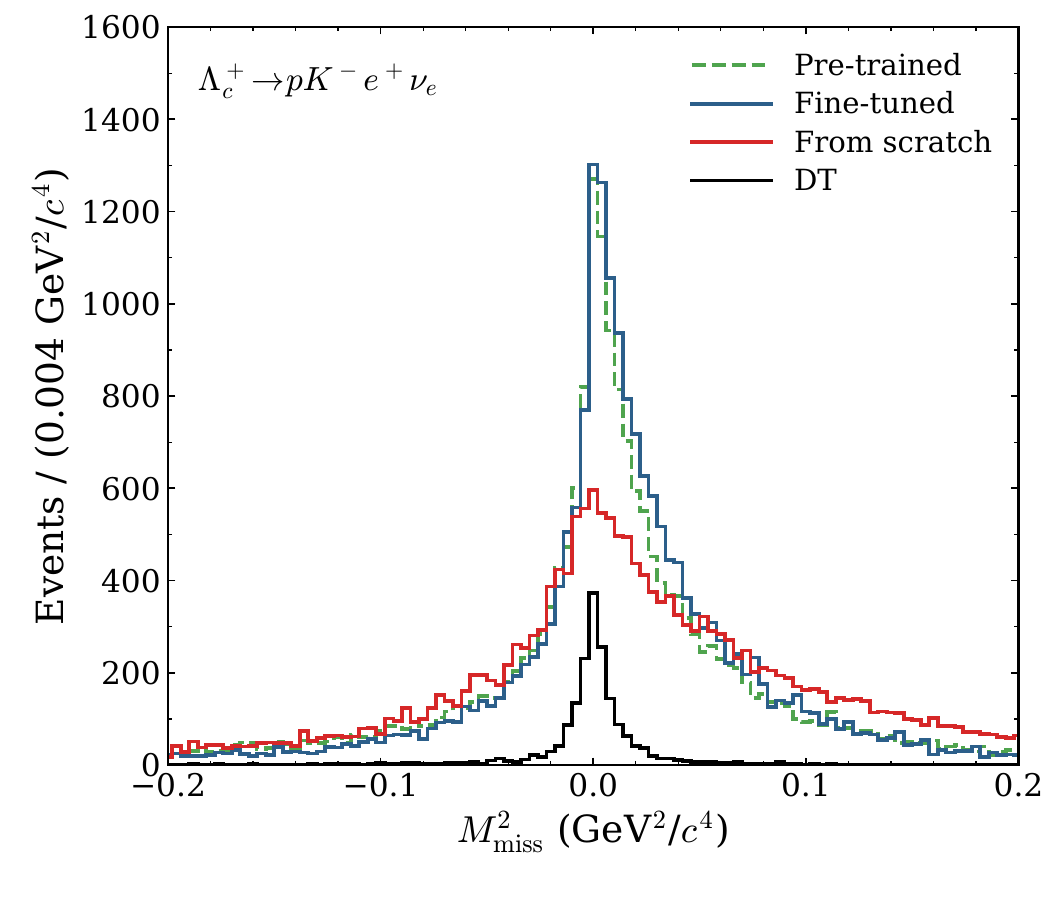}
\caption{The $M_{\rm miss}^{2}$ distributions for simulated $\Lambda_c^+\to pK^-e^+\nu_e$ signal events. The green dashed histogram and the blue, red, and black solid histograms correspond to the pre-trained model, the fine-tuned model, the model trained from scratch, and the DT reference, respectively.}
\label{fig:pkenu-missing-mass}
\end{figure}

Using signal MC events for this decay, we compare a DT reference based on the twelve hadronic tag modes~\cite{12tag,Lc2pKenv} with the three regression models. The $M_{\rm miss}^{2}$ distributions of the four methods are shown in Fig.~\ref{fig:pkenu-missing-mass}. To compare their signal efficiencies under a common selection requirement, the same window $|M_{\rm miss}^{2}|<0.15~\mathrm{GeV}^{2}/c^{4}$ is applied to all four methods~\cite{BESIII:2019DsKenuFF,BESIII:2022DsScalarenu,BESIII:2016Dslnu}, with the corresponding efficiencies listed in Table~\ref{tab:pkenu-window-efficiency}. The pre-trained regression model achieves a signal efficiency $9.63$ times that of the DT reference. Without requiring a recoil-side tag, the fine-tuned model and the model trained from scratch also achieve similarly large gains in signal efficiency, reaching $10.15$ and $9.12$ times that of the DT reference, respectively. Fine-tuning increases the signal efficiency by $11.28\%$ relative to training from scratch. The improved signal efficiency can enhance the sensitivity of searches for rare decays and new physics.

\begin{table}[!htbp]
\centering
\caption{Signal efficiencies of the four methods within the selection
window $|M_{\rm miss}^{2}|<0.15~\mathrm{GeV}^{2}/c^{4}$.}
\label{tab:pkenu-window-efficiency}
\begin{tabular}{lc}
\hline\hline
Method & Signal efficiency (\%) \\
\hline
Double tag & 2.82 \\
From-scratch regression & 25.74 \\
Pre-trained regression & 27.18 \\
Fine-tuned regression & 28.65 \\
\hline\hline
\end{tabular}
\end{table}

\section{Summary}
\label{sec:summary}
Transfer learning strategies based on pre-training and fine-tuning enable the learning of generalizable feature representations from large-scale datasets and facilitate efficient adaptation to downstream tasks.
In this work, 
a general transfer learning strategy is presented for the BESIII experiment. Using $\Lc$ physics as a benchmark, the approach has been demonstrated for both event classification and momentum-direction regression tasks.
Based on the inclusive MC sample covering $\sqrt{s}=4.600$--$4.700~\mathrm{GeV}$,
the pre-trained event classification model rejects $96.90\%$ of the background
for the $\LcLc$ category at a signal efficiency of 90.00\%.
With the training setup adopted in this work, the fine-tuned model outperforms the model trained from scratch across most of the 12 benchmark channels, with particularly pronounced gains in low-statistics regimes.

The pre-trained momentum-direction regression model achieves  average resolutions of
$\sigma(\Delta\theta) = 0.425~\mathrm{rad}$ and
$\sigma(\Delta\phi) = 0.896~\mathrm{rad}$ for inclusive $\Lambda_c^+$ decays.
Across the 12 hadronic benchmark channels, the model provides relative reductions of
$27.34\%$ and $14.11\%$ in the average angular resolutions for $\Delta\theta$ and
$\Delta\phi$, respectively, compared with the conventional ST
reconstruction.
For the representative decay with an undetected neutrino,
$\Lambda_c^+ \to p K^- e^+ \nu_e$, the relative reductions achieved by the
fine-tuned model are $28.39\%$ and $31.69\%$ for the $\Delta\theta$ and
$\Delta\phi$ resolutions, respectively, compared with the model trained from scratch.

In addition, the pre-trained regression model improves the signal efficiency for decays with undetected particles in the final state. For simulated semileptonic decay events within the common selection window, the pre-trained regression model increases the signal efficiency by approximately one order of magnitude relative to the DT reference. Fine-tuning yields an additional improvement over training from scratch, increasing the signal efficiency by $11.28\%$.

This study establishes a scalable and extensible transfer learning paradigm for deep learning at BESIII and provides a foundation for future applications in HEP experiments, such as STCF and CEPC~\cite{STCF, Cheng:2022tog, CEPC}.

\backmatter

\bmhead{Acknowledgements}
The authors thank the BESIII Collaboration for providing the simulation framework, and the IHEP computing center and the supercomputing center of Lanzhou University for their strong support. 
This work is supported in part by National Key R\&D Program of China under Contract No. 2025YFA1613900; CAS Project for Young Scientists in Basic Research under Contract No. YSBR-117; Fundamental Research Funds for the Central Universities; National Natural Science Foundation of China (NSFC) under Contract No. 12221005; the NSFC Excellent Young Scientists Fund Program (Overseas) under Grant No. 2024000542.

\bmhead{Conflict of interest}
The authors declare no conflict of interest.

\bmhead{Data availability statement}
The data used in this study are not publicly available due to collaboration restrictions.

\bmhead{Code availability statement}
This code will be made available on reasonable request.

\bibliography{ref}

\end{document}